\RequirePackage{fix-cm}
\documentclass[smallextended]{svjour3}       
\smartqed  
\usepackage{graphicx}
\usepackage{enumitem}
\usepackage{amsmath}
\usepackage{amsfonts}
\newcommand{\ket}[1]{|{#1}\rangle}
\newcommand{\bra}[1]{\langle{#1}|}

\begin{document}

	\title{From few to many body degrees of freedom \thanks{This work was supported by EPSRC grant No. EP/M024636/1.}
}


\author{Manuel Valiente       
}


\institute{M. Valiente \at
              SUPA, Institute of Photonics and Quantum Sciences,\\
              Heriot-Watt University, Edinburgh EH14 4AS, United Kingdom \\
              \email{M.Valiente\_Cifuentes@hw.ac.uk}           
}

\date{Received: date / Accepted: date}

\maketitle

\begin{abstract}
Here, I focus on the use of microscopic, few-body techniques that are relevant in the many-body problem. These methods can be divided into indirect and direct. In particular, indirect methods are concerned with the simplification of the many-body problem by substituting the full, microscopic interactions by pseudopotentials which are designed to reproduce collisional information at specified energies, or binding energies in the few-body sector. These simplified interactions yield more tractable theories of the many-body problem, and are equivalent to effective field theory of interactions. Direct methods, which so far are most useful in one spatial dimension, have the goal of attacking the many-body problem at once by using few-body information only. Here, I will present non-perturbative direct methods to study one-dimensional fermionic and bosonic gases in one dimension.  
\keywords{Few-body physics \and Scattering theory \and Many-body physics}
\end{abstract}

\section{Introduction}
Strongly interacting quantum many-body ensembles are among the most intriguing and difficult to describe systems in nature. Standard methods such as direct perturbation theory or mean-field approaches are doomed to fail in this regime. Some of these problems, however, can be approximately recast into the form of effective theories which may be either exactly solvable (e.g. integrable quantum systems in one dimension) or weakly-interacting (e.g. dilute Bose gases). The process of obtaining a microscopically accurate effective theory usually involves solving few-body problems, starting with the two-body sector, exactly, in a certain regime of scattering energies or for either weak or strong binding. These constitute what we may call indirect methods: after the simplified theory is obtained, this still needs to be solved, whether exactly or approximately. More recently, renewed interest in the study of strongly interacting one-dimensional systems has arised, especially due to the possibility of preparing and manipulating these using trapped ultracold atoms. In these settings it has been shown recently that few-body solutions can sometimes be used to extract many-body information directly without having to solve a many-body problem at any stage. Examples of direct methods include the recently developed machinery for trapped multicomponent systems, or a few-body method to obtain non-perturbative approximations to the speed of sound in Luttinger liquids. 

\section{Indirect methods}
We assume throughout these notes that we have a non-relativistic many-body system with $N$ particles, interacting via two-body potentials $V$, and possibly under the influence of an external trapping (single-particle) potential $W$. In the first quantisation and in the position representation, the Hamiltonian has the form
\begin{equation}
  H = \sum_{i=1}^N \left[\frac{\mathbf{p}_i^2}{2m}+W(\mathbf{r}_i)\right]+\sum_{i<j=1}^NV(\mathbf{r}_{i,j}),
\end{equation}
where we have assumed, for simplicity, that all particles are identical and have mass $m$, and where $\mathbf{r}_{i,j}\equiv \mathbf{r}_i-\mathbf{r}_j$ are relative coordinates.

We shall call ``indirect'' all those methods that simplify the many-body problem by using few-body techniques. In fact, these methods {\it do not} attempt to solve the many-body problem in any way. We will consider here a number of equivalent methods that aim at simplifying two-body interactions by replacing these with model interactions. This can be achieved by fitting two-body quantities, such as binding energies, scattering data at the relevant scales, or even energy eigenvalues in finite systems or traps, in the following manner: first, the two-body problem with the original, ``realistic'' interaction is solved, and the quantities of interest are determined. Then, the two-body problem with the model, simplified interaction with certain {\it unknown} parameters is solved, and the quantities of interest are fitted to reproduce the ``realistic'' ones. Obviously the first step can be replaced by an experiment if this is available.

\subsection{Expanding the interaction {\it \`a la} EFT}
When applied to interactions, Weinberg's chiral perturbation theory \cite{Weinberg} provides a way to expand the interaction potential between nucleons at low energies. The same principles can be applied in non-relativistic quantum mechanics, following the seminal papers by Phillips {\it et al.} \cite{Phillips} and Kaplan {\it et al.} \cite{Kaplan}. Their expansion is easiest to follow directly in the momentum representation. To do so, we Fourier-transform the two-body interaction potential $V(\mathbf{r})$ ($\mathbf{r}$ is here the relative coordinate), and define
\begin{equation}
  V(\mathbf{k},\mathbf{k}') = \int \mathrm{d}\mathbf{r}e^{i\mathbf{q}\cdot\mathbf{r}}V(\mathbf{r}),
\end{equation}
where $\mathbf{q}\equiv (\mathbf{k}-\mathbf{k}')/2$ is the relative momentum. Even though the interaction potential only depends on $\mathbf{q}$, it is important to keep the $(\mathbf{k},\mathbf{k}')$ dependence in order to avoid confusion at a later stage. We now expand the interaction in a power series of $\mathbf{q}$ (na\"ive expansion)
\begin{equation}
  V(\mathbf{k},\mathbf{k}')=G_0+G_2(\mathbf{k}-\mathbf{k}')^2+\ldots = G_0 + G_2(k^2+k'^2)-2G_2\mathbf{k}\cdot\mathbf{k}'+\ldots .\label{naiveexpansion}
\end{equation}
To $\mathcal{O}(q^2)$, it is clear from Eq.~(\ref{naiveexpansion}) that there is only s-wave and p-wave scattering. It seems, however, too restrictive that the second order s-wave and the lowest order p-wave interaction strengths are proportional to each other, especially if we look at it from a power counting perspective. We shall therefore redefine the expansion, $V_2$, to second order as
\begin{equation}
  V_2(\mathbf{k},\mathbf{k}') = g_0 +g_2(k^2+k'^2)+g_1\mathbf{k}\cdot\mathbf{k}'\equiv V_2^{\mathrm{s}}+V_2^{\mathrm{p}}.
\end{equation}
Fourier-transforming the above interactions back to the position representation yields the pseudo-potential forms in Ref. \cite{Phillips}
\begin{align}
  \bra{\mathbf{r}}V_2^{\mathrm{s}}\ket{\mathbf{r}'}&= \left[g_0 + g_2 (\nabla^2+\nabla'^2)\right]\delta(\mathbf{r}-\mathbf{r}')\delta(\mathbf{r}),\label{naiveS}\\
  \bra{\mathbf{r}}V_2^{\mathrm{p}}\ket{\mathbf{r}'}&= g_1\nabla\cdot\nabla' \delta(\mathbf{r}-\mathbf{r}')\delta(\mathbf{r}).
\end{align}

It is also worth mentioning how the above expansions are done on a lattice, which is a major tool for high-end numerical simulations in a wide variety of fields (for a pedagogical review, see \cite{Nicholson}). We consider a cubic lattice in three dimensions, with lattice spacing $d$. We do not need to pay attention, at this level, to the difference operator representing the kinetic energy, which must have the appropriate continuum limit. To next-to-leading order ($U$-$V$ model), interactions on the lattice are given by
\begin{equation}
  V(\mathbf{n})=U\delta_{\mathbf{n},0}+V\delta_{\mathbf{n},1}+\ldots,
\end{equation}
where $\mathbf{n}=(n_1,n_2,n_3)$ ($n_i\in \mathbb{Z}$) is the relative coordinate, the $d$-dependence is encoded in $U$ and $V$, and $\delta_{\mathbf{n},0}\equiv \prod_{i=1}^3\delta_{n_i,0}$, $\delta_{\mathbf{n},1}\equiv \sum_{\mu=\pm 1}\sum_{i=1}^3\delta_{\mathbf{n},\mu\hat{\mathbf{e}}_i}$ are three-dimensional Kronecker deltas. The (quasi-)momentum representation of the interaction is obtained via the discrete Fourier transform as
\begin{equation}
  V(\mathbf{k},\mathbf{k}')=\frac{1}{d^3}\sum_{\mathbf{n}}V(\mathbf{n})e^{i\mathbf{q}\cdot \mathbf{n} d},
\end{equation}
where $\mathbf{q}=(\mathbf{k}-\mathbf{k}')/2$ and $\mathbf{k},\mathbf{k}'\in (-\pi/d,\pi/d]^3$. The momentum representation of the interaction therefore reads
\begin{align}
  d^3 V(\mathbf{k},\mathbf{k}')&=U+2V\sum_{i=x,y,z}\cos(q_id)\nonumber \\
  &=U+2V\sum_{i=x,y,z}\left[\cos\left(\frac{k_id}{2}\right)\cos\left(\frac{k_i'd}{2}\right)+\sin\left(\frac{k_id}{2}\right)\sin\left(\frac{k_i'd}{2}\right)\right].
\end{align}
The lattice interactions do include, even to this order, infinitely many powers of $\mathbf{q}$. In the continuum limit ($d\to 0$), however, we can adjust $U$ and $V$ such that we reproduce, first, the na{\"i}ve expansion, Eq.~(\ref{naiveexpansion}), by setting
\begin{align}
   &\frac{V}{d^3} = -\frac{4G_2}{d^2}\\
  &\frac{U}{d^3}+3\frac{V}{d^3}=G_0.
\end{align}
We can finally set up the new interaction on the lattice as
\begin{align}
  V_2(\mathbf{k},\mathbf{k}')&=g_0+\frac{12g_2}{d^2}-\frac{8g_2}{d^2}\sum_{i=x,y,z}\cos\left(\frac{k_id}{2}\right)\cos\left(\frac{k_i'd}{2}\right)\nonumber\\
  &+\frac{4g_1}{d^2}\sum_{i=x,y,z}\sin\left(\frac{k_id}{2}\right)\sin\left(\frac{k_i'd}{2}\right).
\end{align}

\subsection{Renormalisation}\label{Renormalisation}
The expansions considered in the previous section should only be regarded as formal series. In any relevant calculation, these potentials lead to complications in the form of ultraviolet (UV) divergences. In some cases these can be easily removed, after regularisation via an UV cutoff $\Lambda$, and the fitting procedure for scattering data or binding energies can be pursued even in the limit $\Lambda\to \infty$ (or in the lattice case, $d\to 0$). It is however now become standard, when the formal limit is not possible, to leave a finite cutoff scale in the theory \cite{Levinsen,Epelbaum1,Epelbaum2}.

The concept of regularisation-renormalisation is easiest exemplified in perturbation theory (see Ref. \cite{Delamotte} for a pedagogical review). We use the lowest-order expansion of the potential, $V_0(\mathbf{k},\mathbf{k}')=g_0$, to fit the effective theory to reproduce the s-wave scattering length $a$ of the original two-body problem. The $T$-matrix is calculated from the Lippmann-Schwinger equation (LSE)
\begin{equation}
  T(z)=V+VG_0(z)T(z),\label{LSE}
\end{equation}
where $G_0(z)$ is the non-interacting Green's function, $z=E+i\eta$, with $E\equiv\hbar^2q^2/m>0$ the scattering energy, and $\eta\to 0^+$. In the momentum representation, the LSE takes the form
\begin{equation}
  \bra{\mathbf{k}}T(z)\ket{\mathbf{k}'} = V(\mathbf{k},\mathbf{k}')+\frac{1}{(2\pi)^3}\int \mathrm{d}\mathbf{s}\frac{V(\mathbf{k},\mathbf{s})\bra{\mathbf{s}}T(z)\ket{\mathbf{k}'}}{E-\hbar^2s^2/m+i\eta},
\end{equation}
Replacing the full interaction $V$ by the model interaction $V_0$ simplifies the above equation, with the $T$-matrix becoming momentum-independent, i.e. $\bra{\mathbf{k}}T(z)\ket{\mathbf{k}'}\equiv t(z) = \mathrm{const.}$. The Born series is easily constructed. The first Born approximation to the $T$-matrix is divergence-free and has the form
\begin{equation}
  t(z) = g_0 +\mathcal{O}(g_0^2).
\end{equation}
At zero energy, the $T$-matrix attains the universal value $t(0)=4\pi\hbar^2a/m\equiv g_R$, which we shall call the renormalised coupling constant -- while $g_0$ is called the bare coupling constant. We expand $g_0$ in series of the renormalised coupling constant as $g_0=g_R+\mathcal{O}(g_R^2)$. Therefore, to lowest order $t(z)=g_R$. The second Born approximation already shows UV-divergent behaviour. It is given by the formal expression
\begin{equation}
  t(z) = g_0+g_0^2 \mathcal{I}(z)+\mathcal{O}(g_0^3),\label{secondBorn}
\end{equation}
where
\begin{equation}
  \mathcal{I}(z)\equiv \frac{1}{(2\pi)^3}\int \mathrm{d}\mathbf{s}\frac{1}{E-\hbar^2s^2/m+i\eta}.
\end{equation}
At zero energy, we obtain, after inserting a spherical cutoff $\Lambda$,
\begin{equation}
  \mathcal{I}(0) = -\frac{m}{2\pi^2\hbar^2}\int_0^{\Lambda} \mathrm{d}s \frac{s^2}{s^2}=-\frac{m\Lambda}{2\pi^2\hbar^2}\to -\infty.
\end{equation}
We now expand the bare coupling constant to second order as $g_0=g_R+Cg_R^2+\mathcal{O}(g_R^3)$. We obtain for the second Born approximation, Eq.~(\ref{secondBorn}), at zero energy
\begin{equation}
  t(0) = g_R+Cg_R^2+g_R^2\mathcal{I}(0) +\mathcal{O}(g_R^2)= g_R +g_R^2\left[C+\mathcal{I}(0)\right]+\mathcal{O}(g_R^3).
\end{equation}
Since, at zero energy, we {\it know} that $t(0)=g_R$, we must set $C=-\mathcal{I}(0)$. Notice that this is exactly equivalent to minimal subtraction. All that is left now is to check the consistency of the method, that is, that setting $C=-\mathcal{I}(0)$ leaves $t(z)$ finite at all energies and that $t(z)$ has predictive power (describes the right physics) at low energies. That $t(z)$ is finite to this order is easily checked as $I(z)-I(0)$ is finite when the limit $\Lambda\to \infty$ is taken. The fact that $t(z)$ describes the right phenomenology at low energy is guaranteed by effective range theory \cite{Taylor}.

It is important to know that other regularisaton schemes lead to the same results after renormalisation. We will illustrate this fact non-perturbatively now. Instead of choosing a spherical cutoff as above, which can be achieved by writing down the interaction as
\begin{equation}
  V_0(\mathbf{k},\mathbf{k}')=g_0\theta(k'^2-\Lambda^2),
\end{equation}
we may choose a ``cubic'' cutoff, more appropriate for lattice applications \cite{ValienteZinner1}
\begin{equation}
  V_0(\mathbf{k},\mathbf{k}')=g_0\theta(\Lambda,\mathbf{k})\theta(\Lambda,\mathbf{k}'),
\end{equation}
where we have defined
\begin{equation}
  \theta(\Lambda,\mathbf{k}) = \prod_{i=x,y,z}\theta(\Lambda^2-k_i^2).
\end{equation}
The $T$-matrix takes the form
\begin{equation}
  \bra{\mathbf{k}}T(z)\ket{\mathbf{k}'}=\frac{\theta(\Lambda,\mathbf{k})\theta(\Lambda,\mathbf{k}')}{1/g_0+\mathcal{I}'_0(z)},
\end{equation}
where
\begin{equation}
  \mathcal{I}'_0(z)=\frac{1}{(2\pi)^3}\int_{\mathbf{s}\in [-\Lambda,\Lambda)^3} \mathrm{d}\mathbf{s}\frac{1}{E-\hbar^2s^2/m+i\eta}.
\end{equation}
The renormalisation prescription $\bra{\mathbf{k}}T(0)\ket{\mathbf{k}'}_{|\mathbf{k}|=|\mathbf{k}'|}=g_R$ is now implemented non-perturbatively, which implies $1/g+\mathcal{I}'_0(0)=1/g_R$, as $\Lambda\to \infty$. Numerically, we obtain $(2\pi)^3\mathcal{I}'_0(0)=-15.3482484734169\ldots \Lambda +\mathcal{O}(1/\Lambda)$. Again, the $T$-matrix is renormalisable if $\mathcal{I}'_0(z)-\mathcal{I}'_0(0)$ remains finite as $\Lambda\to\infty$. To see that, we separate the integral as
\begin{equation}
  (2\pi)^3\mathcal{I}'_0(z)=\int_{s<\Lambda}\mathrm{d}\mathbf{s}\frac{1}{E-\hbar^2s^2/m+i\eta}+\int_{\mathbf{s}\in [-\Lambda,\Lambda)^3 }\mathrm{d}\mathbf{s}\frac{\theta(q^2-\Lambda^2)}{E-\hbar^2s^2/m+i\eta}.\label{I0prime}
\end{equation}
We expand the integrand of the second term in Eq.~(\ref{I0prime}) in a power series which is convergent for all $E>0$, which allows us to rewrite Eq.~(\ref{I0prime}) as
\begin{equation}
  \mathrm{Re}\mathcal{I}'_0(z)=\mathcal{I}'_0(0)-\frac{m}{(2\pi)^3\hbar^2}\sum_{n=1}^{\infty}\int_{\mathbf{s}\in [-\Lambda,\Lambda)^3}\mathrm{d}\mathbf{s}\theta(q^2-\Lambda^2)\frac{k^{2n}}{s^{2(n+1)}},
\end{equation}
while its imaginary part is finite and formally correct. It is not difficult to show that the sum on the second term above vanishes in the limit $\Lambda\to \infty$, which proves renormalisability.
  
\subsection{Pseudopotentials}
So far, we have seen how simplified interactions can be implemented by doing rather na{\"i}ve expansions of the momentum-represented potentials. Their position representations are, on the other hand, a little bit more involved, with Dirac delta functions and off-diagonal elements everywhere. As antinatural as it may appear, the position representation of pseudopotentials came first. In the seminal paper of Huang and Yang \cite{HuangYang}, a family of pseudopotentials was introduced such that they reproduced the low-energy scattering data. In particular, the s-wave Huang-Yang pseudopotential (corresponding to lowest-order interactions) is given by
\begin{equation}
  V_{\mathrm{HY}}(\mathbf{r})=g_R\delta(\mathbf{r})\frac{\partial}{\partial |\mathbf{r}|}(|\mathbf{r}|\cdot).\label{HY}
\end{equation}
Note the difference between $V_{\mathrm{HY}}$ above and the na{\"i}ve bare interaction in Eq.~(\ref{naiveS}). While $g_0\delta(\mathbf{r})$ still needs renormalisation, Eq.~(\ref{HY}) already takes care of it through the differential operator right of the Dirac delta. To see this, let us apply $V_{\mathrm{HY}}$ on a zero-energy wave function of the form $\psi(\mathbf{r})=(1/r-1/a)$. We obtain $V_{\mathrm{HY}}\psi=-(g_R/a)\delta(\mathbf{r})$. The kinetic energy applied on this wave function gives $(4\pi\hbar^2/m)\delta(\mathbf{r})$. In order for $\psi$ to be a solution to the zero-energy stationary Schr{\"o}dinger equation, it is clear that $g_R=4\pi \hbar^2 a/m$, as we already knew from renormalisation.

Much less clear was the situation for momentum-represented pseudopotentials until quite recently. Almost 50 years passed between the appearance of Huang-Yang pseudopotential and the first purely formal definition of its Fourier transform by Tan \cite{Tan1}, which made it difficult to reconcile standard renormalisation methods with the ``renormalisation-by-differentiation'' approach of Huang and Yang. In fact, Tan's momentum representation of the Huang-Yang pseudopotential was considered bizarre for a number of years \cite{BraatenPhysics,Combescot}, since it appeared as if he had introduced some novel type of mathematical distributions. This is not the case, as we will see below.

Tan defined the action of Huang-Yang pseudopotential on a plane wave as
\begin{equation}
  \delta(\mathbf{r})\frac{\partial}{\partial r}\left(re^{i\mathbf{k}\cdot\mathbf{r}}\right)\equiv \delta(\mathbf{r}) \Lambda(\mathbf{k}).\label{Lambdadistro}
\end{equation}
Above, $\Lambda$ obviously satisfies $\Lambda(\mathbf{k})=1$ if $k<\infty$, but also satisfies
\begin{equation}
  \int \mathrm{d}\mathbf{k}k^{-2}\Lambda({\mathbf{k}})=0.
\end{equation}
The above integral relations can be proved following Ref. \cite{Tan1}, by multiplying Eq.~(\ref{Lambdadistro}) by $k^{-2}$ and integrating over $\mathbf{k}$. Knowing these properties, it is actually very easy to obtain an explicit form for $\Lambda(\mathbf{k})$, using only well-known distributions. This was shown in Ref. \cite{ValienteTan}, and the $\Lambda$-distribution takes the form
\begin{equation}
  \Lambda(\mathbf{k})=1-\frac{\delta(1/k)}{k}.\label{FourierTransform}
\end{equation}

We show now that the Fourier-transformed pseudopotential, Eq.~(\ref{FourierTransform}), multiplied by the renormalised coupling constant $g_R$, yields the correct solution to the two-body problem. In this case, we look at the bound state problem. The integral equation is given by
\begin{equation}
  1=-\frac{g_R}{2\pi^2}\int_0^{\infty}\mathrm{d}k\frac{k^2\Lambda(k)}{\hbar^2k^2/m+E_B},\label{equbound}
\end{equation}
where $E_B=-E>0$ is the binding energy to be calculated. We rewrite the integral as
\begin{equation}
  \int_0^{\infty}\mathbf{k}\frac{k^2\Lambda(k)}{\hbar^2k^2/m+E_B}=\lim_{K\to \infty}\left[\int_{0}^K\mathbf{d}k\frac{k^2}{\hbar^2k^2/\mu+E_B}-\int_{1/K}^{\infty}\mathrm{d}\tau \frac{\delta(\tau-1/K)\tau^{-1}}{\hbar^2/\mu+E_B\tau^2}\right]^{-1},
\end{equation}
the limit above being set by definition, and where we have changed variables as $k=1/\tau$. Minor algebraic manipulations give the solution $E_B=\hbar^2/ma^2$ to Eq.~(\ref{equbound}), as should be.

\subsection{L{\"u}scher's method}
We now move on to the last method to extract low-energy interactions. Part of the method is due to L{\"u}scher \cite{Luscher}, whose motivation (among other) was to extract scattering phase shifts of either realistic two-body interactions or emergent interactions between composite particles. Nowadays L{\"u}scher's results and generalisations thereof are used as a powerful tool to derive effective or model interactions at the few-body level that simplify the theoretical or computational analysis of systems with higher numbers of particles.

The method works essentially as follows. We place the system of interest in a finite volume, and choose some boundary conditions (periodic, open, etc.). In the simplest case with no bound states present, the lowest eigenenergy of the system no longer vanishes, but is either lifted or lowered depending on the properties of the interaction. If we are interested in the low-energy properties of the system, we can focus on the ground state energy and its volume dependence, which we denote $E_0(\Omega)$, with $\Omega$ the volume. In the case of s-wave scattering, $E_0(\Omega)$ will depend on the scattering length $a$ and the effective range $r$ to leading and next-to-leading orders in powers of inverse volume. We can therefore diagonalise the two-body system in a finite volume with a generic low-energy effective interaction for different values of $\Omega$, and extract the interdependence of the interaction parameters with the scattering length and effective range, and once these agree with the physical values, we have fitted the model. Below, we present the simplest example of lowest order s-wave interactions using effective field theory \cite{BeaneLuscher,ValienteZinner1}.

We begin by writing the stationary Schr{\"o}dinger equation for the Hamiltonian $H=H_0+V$, after separation of centre of mass and relative coordinates with periodic boundary conditions (in the relative coordinate). Since the system is finite, the spectrum of $H$ is discrete and even its positive eigenvalues are different from those of $H_0$, a solution $\ket{\psi}$ to the Schr{\"o}dinger equation with energy $E$ satisfies
\begin{equation}
  \ket{\psi}=G_0(E)V\ket{\psi},\label{LSEbound}
\end{equation}
where $G_0(E)=(E-H_0)^{-1}$ is the non-interacting Green's function. It is straightforward to show that Eq.~(\ref{LSEbound}) can be rewritten as
\begin{equation}
  \ket{\psi}=\mathbf{k}+\left[G_0(E)V-\ket{\mathbf{k}}\bra{\mathbf{k}}G_0(E)V\right]\ket{\psi},\label{BG}
\end{equation}
with the supplementary condition $E=\hbar^2k^2/m+\bra{\mathbf{k}}V\ket{\psi}$ \cite{ValienteZinner1}. In Eq.~(\ref{BG}), $\mathbf{k}$ is a {\it normalised} non-interacting eigenstate of the Hamiltonian with energy $\hbar^2k^2/m$. Defining $V\ket{\psi}=\hat{r}\ket{\mathbf{k}}$, and using Eq.~(\ref{BG}), we obtain
\begin{equation}
  \bra{\mathbf{k}'}\hat{r}\ket{\mathbf{k}}=\bra{\mathbf{k}'}V\ket{\mathbf{k}}+\sum_{\mathbf{q}\ne \mathbf{k}} \frac{\bra{\mathbf{k}'}V\ket{\mathbf{q}}\bra{\mathbf{q}}\hat{r}\ket{\mathbf{k}}}{E-\hbar^2q^2/m},\label{BGE}
\end{equation}
while the auxiliary condition reads $E=\hbar^2k^2/m+\bra{\mathbf{k}}\hat{r}\ket{\mathbf{k}}$. If we particularise to the lowest order s-wave interaction with bare coupling constant $g_0$, Eq.~(\ref{BGE}) becomes very simple and the solution is a constant ($ \bra{\mathbf{k}'}\hat{r}\ket{\mathbf{k}}=r(E)$), given by
\begin{equation}
  r(E)=\frac{1}{1/g_0-\sum_{\mathbf{q}\ne\mathbf{k}}[E-\hbar^2q^2/m]^{-1}}.
\end{equation}
Using the above equation and the auxiliary condition for the energy yields the final non-linear equation
\begin{equation}
  E = \frac{\Omega^{-1}}{1/g_0-\Omega^{-1}\sum_{\mathbf{q}\ne 0}[E-\hbar^2q^2/m]^{-1}},\label{Luscher1}
\end{equation}
where we have chosen $\mathbf{k}=0$ without loss of generality. The sum in Eq.~(\ref{Luscher1}) is UV-divergent. Like in the continuum, integral case, there are several equivalent ways to regularise the sum. The spherical cutoff scheme has been used by Beane {\it et al.} \cite{BeaneLuscher}. However, it seems more natural to use a cubic cutoff scheme here due to the periodic boundary conditions. Expressing the volume as $\Omega=L^3$, and explicitly writing the quantised momenta as $\mathbf{q}=2\pi\mathbf{n}/L$, with $\mathbf{n}\in \mathbb{Z}^3$, the regularised sum takes the form
\begin{equation}
  S_{\lambda}(E)=\frac{1}{L^3}\sum_{\mathbf{n}\ne 0}\frac{\theta(\mathbf{n},\lambda)}{E-\hbar^2(2\pi |\mathbf{n}|/L)^2/m},
\end{equation}
where $\lambda$ is an integer cutoff such that $\Lambda=2\pi \lambda/L$. The renormalisation of the interaction was given in Section \ref{Renormalisation}, as $1/g_0 = 1/g_R-\mathcal{I}_0'(0)$. Therefore, Eq.~(\ref{Luscher1}) becomes
\begin{equation}
  E=\frac{L^{-3}}{1/g_R-s(E)},\label{Luscher2}
\end{equation}
where we have defined the renormalised sum $s(E)=S_{\lambda}(E)+\mathcal{I}_0'(0)$. For small scattering lengths, we can expand Eq.~(\ref{Luscher2}) in powers of $g_R$. To next-to-leading order, we obtain
\begin{equation}
  E=\frac{g_R}{L^3}+\frac{g_R^2}{L^3}s(0)+\mathcal{O}(g_R^3).
\end{equation}
The renormalised sum can be evaluated numerically, and after using $g_R=4\pi \hbar^2 a/m$, we obtain
\begin{equation}
  E=\frac{4\pi\hbar^2}{m}\frac{a}{L^3}\left[1+2.837297\frac{a}{L}\right] +\mathcal{O}(g_R^3),
\end{equation}
in agreement with L{\"u}scher's analysis \cite{Luscher}.

\section{Application of indirect methods in the many-body problem}
We now focus our attention on a simple application of indirect methods to the many-body problem. For simplicity, we consider ultracold bosons with s-wave interactions. We will make use of perturbative renormalisation in bosonic mean-field theory and Bogoliubov theory.

\subsection{Mean-field theory}
We consider a many-boson system interacting via pairwise potentials $V$. We assume that the system is dilute, that is, the density $\rho$ and the s-wave scattering length satisfy $0<\rho a^{1/3}\ll 1$. We may directly write down the simplified Hamiltonian including the bare, lowest-order s-wave interaction, which reads
\begin{equation}
  H=\sum_{\mathbf{k}}\epsilon(\mathbf{k})b_{\mathbf{k}}^{\dagger}b_{\mathbf{k}}+\frac{g_0}{2\Omega}\sum_{\mathbf{k}\mathbf{k}'\mathbf{q}}b_{\mathbf{k}+\mathbf{q}}^{\dagger}b_{\mathbf{k}'-\mathbf{q}}^{\dagger}b_{\mathbf{k}'}b_{\mathbf{k}},
\end{equation}
where $\epsilon(\mathbf{k})=\hbar^2k^2/2m$, and $b_{\mathbf{k}}^{\dagger}$ ($b_{\mathbf{k}}$) is the bosonic creation (annihilation) operator with momentum $\mathbf{k}$.

Lowest-order mean-field theory corresponds to expanding the bare interaction to lowest order, which as we know means $g_0=g_R+\mathcal{O}(g_R^2)$, while using the non-interacting ground-state as an ansatz. The expectation value of the Hamiltonian gives an estimate of the energy $\langle H \rangle = g_RN(N-1)/(2\Omega)$ which, in the thermodynamic limit, gives the mean-field energy $E_{\mathrm{MF}}$ 
\begin{equation}
\frac{E_{\mathrm{MF}}}{N}=\frac{g_R\rho}{2}.
\end{equation}

In order to go beyond the lowest-order approximation, we need to invoke Bogoliubov's theory (see \cite{Fetter}). All that is relevant for our purposes is the final expression for the ground-state energy $E_{\mathrm{Bog}}$, which has the form, in the thermodynamic limit
\begin{equation}
  \frac{E_{\mathrm{Bog}}}{N}=\frac{g_0\rho}{2}-\frac{1}{2\rho}\frac{1}{(2\pi)^3}\int \mathrm{d}\mathbf{k}\left[\epsilon(\mathbf{k})+g_0\rho-\mathcal{E}(\mathbf{k})\right].\label{BogoEnergy}
\end{equation}
Above, $\mathcal{E}(\mathbf{k})$ is the quasi-particle energy, given by
\begin{equation}
  \mathcal{E}(\mathbf{k})=\sqrt{2g_0\rho\epsilon(\mathbf{k})+[\epsilon(\mathbf{k})]^2}.\label{qpar}
\end{equation}
To see that the ground-state energy (\ref{BogoEnergy}) is renormalised to second order in the coupling constant, we need to look into the UV-behaviour of the integrand in Eq.~(\ref{BogoEnergy}). The quasi-particle energy, Eq.~(\ref{qpar}), is expanded in series of the bare coupling constant. This yields
\begin{equation}
  \mathcal{E}(\mathbf{k})=\epsilon(\mathbf{k})+g_0\rho-\frac{g_0^2\rho^2}{2\epsilon(\mathbf{k})}+\ldots
\end{equation}
Notice that all terms omitted above yield integrals over $\mathbf{k}$ that are convergent in the UV.
Given that $g_0=g_R+Cg_R^2+\mathcal{O}(g_R^3)$ with $C=-\int \mathrm{d}\mathbf{k}(1/2\epsilon(\mathbf{k}))$ in {\it any} regularisation-renormalisation scheme, we see that the ground-state energy is convergent and given by
\begin{equation}
  \frac{E_{\mathrm{Bog}}}{N}=\frac{g_R\rho}{2}-\frac{1}{2\rho}\frac{1}{(2\pi)^3}\int \mathrm{d}\mathbf{k} \left[\epsilon(\mathbf{k})+g_R\rho-\frac{g_R^2\rho^2}{2\epsilon(\mathbf{k})}-\mathcal{E}_R(\mathbf{k})\right],
\end{equation}
where $\mathcal{E}_R(\mathbf{k})$ is given by Eq.~(\ref{qpar}) with $g_0$ replaced with $g_R$.

\subsection{Luttinger liquids}
One-dimensional many-body systems are markedly different from their three-dimensional counterparts. The usual 3D paradigms for bosons and fermions, especially in the weakly-interacting or weak-coupling limit, are well known. A cold Bose gas in the dilute limit can be described by means of mean-field and Bogoliubov theories. The excitations -- phonons -- are linearly dispersed and gapless. In the fermionic case, standard perturbation theory can be used, and the quasi-particles are dressed fermions, not very different from the bare ones, and are essentially non-interacting with a different effective mass. In the 1D case, both weakly and strongly interacting systems, when these are gapless, are described at low energies in a unified manner by Luttinger liquid theory. In essence, the original Hamiltonian of the system, which need not be exactly solvable, is replaced by an effective Hamiltonian (Tomonaga-Luttinger Hamiltonian \cite{Bruus}), which can be solved analytically via bosonisation \cite{Mattis}. Below we show how this comes about for spinless fermions using the so-called ``constructive bosonisation'', which has a strong few-body component. We note that, for bosons, it is best to consider field-theoretic bosonisation as described by Haldane \cite{Haldane}.

The main idea behind Luttinger liquid theory is the fact that non-interacting fermions near the Fermi points have a linear dispersion. Excitations around the Fermi points ($\pm k_F$) have the following frequency
\begin{equation}
\hbar \omega (q) = \frac{\hbar^2}{2m}\left[(\pm k_F\pm q)^2-k_F^2\right]=\hbar v_Fq+\frac{\hbar^2q^2}{2m}, 
\end{equation}
where we have defined the Fermi velocity $v_F= \hbar k_F/m$. The above full dispersion of excitations is simplified to $\hbar \omega(q)= \hbar v_F q$ in Luttinger liquid theory. This is equivalent to setting, as a {\it single-particle} dispersion
\begin{equation}
  \epsilon(k) = E_F+\hbar v_F(|k|-k_F).
\end{equation}
The other usual procedure in Luttinger liquid theory has to do with the two-body interaction. In this case, it is not low-energy collisions that really matter in the problem, since we are interested in physics around the Fermi points. Moreover, interactions between two fermions near the same Fermi point is strongly suppressed by the Pauli principle. We shall therefore focus on interactions between two fermions at opposite Fermi points. The energy of the collision between these two particles is just $2E_F=\hbar^2k_F^2/m$. Since we have fermions, we may symmetrise the interaction. Any two-body potential $V(k,k')\equiv V(q)$ can be written as
\begin{equation}
  V(k,k')=V_s(k,k')+V_p(k,k'),
\end{equation}
where $s$ and $p$ refer to even and odd wave, respectively, and are given by
\begin{align}
  V_s(k,k')&=\frac{1}{2}\left[V(k,k')+V(k,-k')\right],\\
  V_p(k,k')&=\frac{1}{2}\left[V(k,k')-V(k,-k')\right].\label{Vp}
\end{align}
The odd wave interaction is the only one having an effect on fermion-fermion scattering. The value of the interaction between fermions at opposite Fermi points is given by
\begin{equation}
V_p(\pm k_F,\pm k_F)=\frac{1}{2}\left[V(k_F,-k_F)-V(k_F,k_F)\right]=\frac{1}{2}\left[V(0)-V(2k_F)\right].\label{Vpconst}
\end{equation}
In the condensed-matter community, the whole two-body interaction is typically replaced by the above constant, which is obviously wrong (they are fermions), and leads to puzzling and trivial contradictions that are later resolved by other physical arguments \cite{Bruus}. This issue is not present at all from the beginning if we simply recall that Eq.~(\ref{Vpconst}) comes from an odd wave interaction, Eq.~(\ref{Vp}), which satisfies $V_p(k,-k')=-V_p(k,k')$. The simplest possible interaction that satisfies the two requirements is therefore given by
\begin{equation}
  V_p(k,k')\approx\frac{1}{2}\left[V(0)-V(2k_F)\right]\mathrm{sgn}(k)\mathrm{sgn}(k').
\end{equation}
As usual, the strength of the effective interaction obtained in this way is just the first Born approximation, and we must replace the above value by a bare coupling constant $g_0$, and define the effective interaction $V_0$ as
\begin{equation}
  V_0(k,k')=g_0\mathrm{sgn}(k)\mathrm{sgn}(k').
\end{equation}  
The scattering states in the position representation, before linearising the dispersion, can be calculated easily by solving the Lippmann-Schwinger equation in the momentum representation and Fourier-transforming the corresponding scattering states. Asymptotically, these take the form
\begin{equation}
  \psi_k(x)\propto \sin(kx)-\frac{mg_0}{2\hbar^2k}\mathrm{sgn}(x)\cos(kx),
\end{equation}
where $x=x_1-x_2$ is the relative coordinate. Since in 1D two-fermion scattering states have the asymptotic behaviour $\psi_k(x)\propto \mathrm{sgn}(x)\sin(k|x|+\theta_k)$, with $\theta_k$ the scattering phase shift, we see that the coupling constant must take the value
\begin{equation}
  g_0=-2\hbar v_F \tan \theta_{k_F},\label{g0value}
\end{equation}
in order to reproduce the scattering phase shift $\theta_{k_F}$

The situation is quite different if we linearise the dispersion $\epsilon(k)$. In this case, the inverse of the real part $\mathcal{T}$ of the on-shell $T$-matrix is UV-divergent, and takes the form $\mathcal{T}=\tau\mathrm{sgn}(k)\mathrm{sgn}(k')$, with
\begin{equation}
  \tau = \frac{1}{\frac{1}{g_0}-\frac{1}{2\pi\hbar v_F}\log\left|\frac{\Lambda-|k|}{|k|}\right|}.\label{tau}
\end{equation}
The way to sort this divergence out is to realise that we have done two simplifications: (1) we have substituted the full interaction by a model one and (2) we have linearised the dispersion. We then have two degrees of freedom to fit -- or renormalise -- the theory to. In this way, we can renormalise as $g_0=-2\hbar v_F \tan \theta_{k_F}$ (Eq.~(\ref{g0value})), and regard the Fermi point $k_F$ as a {\it bare} Fermi momentum, by replacing it with $k_F\to k_F^{(0)}$. Since we are interested in relative momenta around the (bare) Fermi points, as their location is unimportant, we may set $k_F^{(0)}=\Lambda/2+k_F$, which renders Eq.~(\ref{tau}) finite around the bare Fermi points, while it suppresses scattering at irrelevant momenta deep in the Fermi sea. This procedure leads in a more natural way to Luttinger's model \cite{Luttinger}, where the system is separated into right and left movers whose momenta can in both cases take any value in $(-\infty,\infty)$. To arrive at Luttinger's model, we shift the momentum of right (left) movers as $k\to k+\Lambda/2$ ($k-\Lambda/2$), and rescale the cutoff as $\Lambda/2\to \Lambda$. In this way, the momenta of right and left movers both belong to the interval $(-\Lambda,\Lambda)$. The formal limit $\Lambda\to \infty$ can be taken and the non-interacting Hamiltonian becomes (ignoring constant terms)
\begin{equation}
  H_0=\hbar v_F\sum_{k=-\infty}^{\infty}\left[c_{kR}^{\dagger}kc_{kR}-c_{kL}^{\dagger}kc_{kL}\right],\label{LuttingerHam}
\end{equation}
where we have defined $c_{kR}$ ($c_{kL}$) as the annihilation operator for a right (left) mover with momentum $k+\Lambda$ ($k-\Lambda$). Hamiltonian (\ref{LuttingerHam}) corresponds exactly with the kinetic energy term in Luttinger's Hamiltonian. We note that the main difference between Luttinger's and Tomonaga-Luttinger's models is that the former is exactly solvable even in the interacting case and, as we have seen, poses no problems of infinities.

The bosonisation solution to Luttinger's model can be found in many texts and reviews in the literature. See, for instance Ref. \cite{Pedestrian}. The solution unveils collective bosonic excitations that are linearly dispersed, i.e. $\hbar \omega (q) = \hbar v q$, where the speed of sound is given by
\begin{equation}
  v = v_F\sqrt{1+\frac{2g_0}{\pi \hbar v_F}}.
\end{equation}
The exact value of $g_0$ must in principle be calculated fully numerically. However, we do have a non-perturbative estimate for it from our two-body analysis, Eq.~(\ref{g0value}). In the next section we will consider a much more accurate method to extract the value of the speed of sound.


\section{Direct methods}
We shall call ``direct'' those methods that attempt to describe properties of a many-body system using few-body techniques or information. These methods, so far, have mostly succeeded in one spatial dimension, due to its simplicity, and we will describe a number of these one-dimensional systems. Perhaps the most famous and important method that may be considered direct is the Bethe ansatz, which has been the subject of intense investigation (for a pedagogical review see \cite{Bethe}), and will not be considered here. The last of the subsections, dealing with flat bands, can be applied to any dimension, but is only discussed for the one dimensional case.

\subsection{Strongly-interacting multicomponent systems in one dimension}
We consider a one-dimensional system of particles interacting via Dirac delta potentials under the influence of an external trap. The Hamiltonian reads
\begin{equation}
  H=-\frac{\hbar^2}{2m}\sum_{i=1}^N\frac{\partial^2}{\partial x_i^2} +\sum_{i=1}^NW(x_i)+g\sum_{i<j=1}^N\delta(x_i-x_j).\label{hamiltonian1D}
\end{equation}
For concreteness, we will focus on spin-$1/2$ fermions, with $N_{\uparrow}$ spin-up and $N_{\downarrow}$ spin-down fermions. The homogeneous case ($W=0$) is exactly solvable via the Bethe ansatz for all $g$ \cite{YangBethe}. Much less is known about the trapped case, which is nevertheless the physically relevant case in cold atom experiments \cite{Blochreview,Selim1}. For arbitrary interaction strength $g$, numerical results do exist in the literature (see, e.g. Ref. \cite{Lindgren}). However, complexity quickly rises with particle number, and other more analytical approaches are important. In the strongly-interacting limit, $g\to \infty$, both the ground state and some excitations can be accessed using a method first described in Ref. (\cite{Volosniev}).

The idea goes as follows. In the strongly interacting regime, two bosons (or, equivalently two spin-$1/2$ fermions in a singlet state) will ``fermionise'' when the zero-range interaction between them becomes infinitely large \cite{Girardeau}. That is, two-boson scattering states of Hamiltonian (\ref{hamiltonian1D}) take the form
\begin{align}
  \Psi_B(x_1,x_2)=e^{iK(x_1+x_2)/2}\sin(k|x_1-x_2|)&=\mathrm{sgn}(x_1-x_2)e^{iK(x_1+x_2)/2}\sin(k(x_1-x_2))\nonumber \\
  &=\mathrm{sgn}(x_1-x_2)\Psi_F(x_1,x_2),
\end{align}
where $\Psi_F$ is a two-body non-interacting state of spin-polarised (or spin-triplet) fermions. Moreover, in a finite many-boson system or in the thermodynamic limit, bosonic eigenstates can be obtained trivially from their spin-polarised, non-interacting fermionic counterparts as $\Psi_B(x_1,\ldots,x_N)=\prod \left[\mathrm{sgn}(x_i-x_j)\right]\Psi_F(x_1,\ldots,x_N)$, and their ground state energies and excitations coincide \cite{Girardeau}. The mapping just described -- Bose-Fermi mapping -- is also valid when external traps are present. Lowest order corrections (in $1/g$) to the ground state energy are easily obtained by means of the tails of the momentum distributions \cite{Olshanii,ValienteEquiv}.

The situation is more complicated for multicomponent and spinful systems. Girardeau's fermionised wave functions are still exact eigenstates in the strict $1/g=0$ limit. However, the states are highly degenerate and no single eigenstate connects adiabatically to Girardeau's state from finite $1/g$ \cite{Volosniev}. What remains true for spinful systems is that the eigenstates are {\it piecewise} fermionised. We can write a general state as
\begin{equation}
  \Psi(x_1,\ldots,x_N)=\sum_{k=1}^{N!}a_k\theta(x_{P_k(1)},\ldots,x_{P_k(N)})\Psi_F(x_1,\ldots,x_N).\label{generalstate}
\end{equation}
Above, the sum is over the $N!$ permutations of the $N$ positions, and $\theta(x_1,\ldots,x_N)=1$ for $x_1<x_2<\ldots<x_N$, and vanishes otherwise. Due to symmetry, for spin-$1/2$ fermions there are only $N!/(N_{\downarrow}!N_{\uparrow}!)$ physically different permutations (or sectors), which reduces the number of independent coefficients $a_k$. To visualise this take $N=4$ particles. Particles labelled as $1$ and $2$ have spin up and those labelled as $3$ and $4$ have spin down. Denoting configurations $x_{1}<x_{2}<x_{3}<x_{4}$ as $\uparrow\uparrow\downarrow\downarrow$, since $1$ and $2$ are identical and so are $3$ and $4$, we see that the number of degenerate states (and therefore inequivalent configurations) is $6$, by simply enumerating all possible configurations of two $\uparrow$'s and two $\downarrow$'s. We now use the state $\Psi$ of Eq.~(\ref{generalstate}) as a variational ansatz. The perturbed energy away from $1/g$ is given by $E=E_0-K/g$, where $E_0$ is the fermionised energy (the sum of all first $N$ single-particle eigenenergies of the non-interacting Hamiltonian), and we use the Hellmann-Feynmann theorem to show that
\begin{equation}
  K=\lim_{g\to \infty}g^2\frac{\sum_{i<j}\int \mathrm{d}\mathbf{X}\left|\Psi\right|^2\delta(x_i-x_j)}{||\Psi||^2}.\label{equ1}
\end{equation}
Above, we have defined $\mathrm{d}\mathbf{X}=\prod \mathrm{d}x_i$.
We now use the boundary conditions implied by the Dirac delta interaction
\begin{equation}
  \left(\frac{\partial \Psi}{\partial x_i}-\frac{\partial \Psi}{\partial x_j}\right)_{x_{i}-x_j\to 0^+}-\left(\frac{\partial \Psi}{\partial x_i}-\frac{\partial \Psi}{\partial x_j}\right)_{x_{i}-x_j\to 0^-}=2g\Psi(x_i=x_j),\label{equ2}
\end{equation}
in Eq.~(\ref{equ1}), make use of the equivalency of identical configurations, and the explicit form of the wave function $\Psi$ to write
\begin{equation}
  K=\frac{\sum_{n\ne m}(a_n-a_m)^2\alpha_{n,m}}{\sum_na_n^2},\label{K}
\end{equation}
with
\begin{equation}
  \alpha_{n,m}=\int_{\Gamma_n}\mathrm{d}\mathbf{X}\delta(x_i-x_j)\left|\left(\frac{\partial\Psi_F}{\partial x_i}\right)\right|^2,
\end{equation}
where $\Gamma_n$ is the region with $x_{P_n(1)}<x_{P_n(2)}<\ldots<x_{P_n(N)}$, with $P_n$ is a permutation such that $x_i$ and $x_j$ are adjacent to each other, while $P_m$ denotes the same permutation with $x_i$ and $x_j$ in reverse order.

It is perhaps easier to digest the above theory by considering a three-body system with $N_{\uparrow}=2$. The three different configurations are given by $\ket{1}=\ket{\downarrow\uparrow\uparrow}$, $\ket{2}=\ket{\uparrow\downarrow\uparrow}$ and $\ket{3}=\ket{\uparrow\uparrow\downarrow}$. Therefore, there are three different coefficients $(a_1,a_2,a_3)$. The energy functional takes the form
\begin{equation}
  (\sum_{i=1}^3a_i^2)K=(a_1-a_2)^2\alpha_{1,2}+(a_1-a_3)^2\alpha_{1,3}+(a_2-a_3)^2\alpha_{2,3}.
\end{equation}
The $\alpha_{i,j}$ are given by
\begin{align}
  \alpha_{1,2}&=\int_{\downarrow\uparrow\uparrow}\mathrm{d}x_1\mathrm{d}x_2\mathrm{d}x_3\delta(x_1-x_2)\left|\partial_{x_1}\Psi_F\right|^2,\\
  \alpha_{1,3}&=\int_{\downarrow\uparrow\uparrow}\mathrm{d}x_1\mathrm{d}x_2\mathrm{d}x_3\delta(x_1-x_3)\left|\partial_{x_1}\Psi_F\right|^2=0,\\
  \alpha_{2,3}&=\int_{\uparrow\downarrow\uparrow}\mathrm{d}x_1\mathrm{d}x_2\mathrm{d}x_3\delta(x_2-x_3)\left|\partial_{x_2}\Psi_F\right|^2.
\end{align}

The energy functional $K$ in Eq.~(\ref{K}) can be used to extract all the eigenstates at $g=\infty$ and the corrections to their energies to order $1/g$. To do so, we take the functional (where we re-instate the {\it a priori} complex nature of the coefficients)
\begin{equation}
  \mathcal{F}=\sum \alpha_{i,j}|a_i-a_j|^2-\kappa\sum_i|a_i|^2,
\end{equation}
where $\kappa$ is a Lagrange multiplier that is identified with $K$ after variation ($\kappa=K$). By optimising $\delta \mathcal{F}/\delta a_k^*=0$, we obtain an eigenvalue equation. For the three-body case discussed above, we obtain
\begin{align}
  &(a_1-a_2)\alpha_{1,2}=Ka_1,\\
  -&(a_1-a_2)\alpha_{1,2}+(a_2-a_3)\alpha_{2,3}=Ka_2,\\
  -&(a_2-a_3)\alpha_{2,3}=Ka_3.
\end{align}

\subsection{Luttinger liquids}
We now discuss a method that attacks the problem of calculating the speed of sound of a Luttinger liquid directly using two-body physics \cite{ValienteFew}. 

We begin by considering two non-relativistic spinless fermions in a box of length $L$ under periodic boundary conditions. We would like to look at physics near a hypothetical Fermi sea, using a finite two-body system. The most relevant process in the many-body system is the scattering of particles with momenta at opposite Fermi points $\pm k_F$. We therefore set the conserved centre of mass momentum $K=0$. In the non-interacting case, the single-particle momenta are quantised as $k=2\pi n/L$. Therefore, the lowest-energy state without interactions is given in the first quantisation by
\begin{equation}
  \ket{0}=\frac{1}{\sqrt{2}}\left(\ket{-k_F,k_F}-\ket{k_F,-k_F}\right),
\end{equation}
where $k_F\equiv 2\pi/L$. Lowest order in perturbation theory gives the following energy $E_1(L)$, which is size dependent
\begin{equation}
  E_1(L)=\frac{\hbar^2k_F^2}{m}+\frac{1}{L}\left[V(0)-V(2k_F)\right]=2E_F+\frac{2}{L}V_p(k_F,k_F).\label{perten}
\end{equation}
Above, $V_p(k_F,k_F)$ should already be familiar from the previous analysis of Luttinger liquids. In Eq.~(\ref{perten}) we may already identify the density $\rho = k_F/\pi=2/L$.

If we are to study low-lying excitations above the ground state using a finite two-body system, we encounter the difficulty of quantisation. The lowest lying excitations, before interactions are included, correspond to taking one particle from the Fermi points $\pm k_F$ and promoting it to $\pm 2k_F$. The momentum of the excitation is therefore $q=k_F=2\pi/L$, far too large to be considered low-lying. To circumvent this issue, we may instead shrink the length of the system. We place the two particles on a ring of length $L-\delta$, with $0<\delta\ll L$, and expand $(L-\delta)^{-1}$ to lowest order in $\delta$. We obtain
\begin{equation}
  \frac{1}{L-\delta}=\frac{1}{L}+\frac{1}{L^2}\delta+\mathcal{O}(\delta^2)\equiv\frac{1}{2\pi}\left( k_F+\frac{q}{2}\right).
\end{equation}
Above, we have identified $q$ with the excitation momentum, since both particles are excited by an amount $\delta/L^2$, amounting to two excitations of magnitude $q/2$ each. Notice that by shrinking the length of the system by an infinitesimal amount $\delta$ we have solved the problem of large momenta. However, a new problem has also arised: by shrinking the size of the system we have also increased the density. The only way of dealing with this problem is by recalling the fact that in a thermodynamically large system (our target system) it is not the interaction strength nor the density alone that defines the properties of the system. These are in fact defined by the {\it dimensionless} coupling constants of the many-body system. For example, lowest order EFT for fermions in 1D corresponds to the potential $V(q)=g_2q^2$. To lowest order in perturbation theory, $g_2=-(2\hbar^2/mk_F)\tan\theta_{k_F}$ (coinciding with the renormalised coupling constant). The only dimensionless parameter of the theory at finite density $\rho=k_F/\pi$ is the so-called Lieb-Liniger parameter $\gamma$, given by
\begin{equation}
  \gamma=-\frac{4\pi\hbar^2}{mk_F}\frac{1}{g_2}.\label{gamma}
\end{equation}
Therefore, in order for the two-body system to describe the same physics in the boxes of sizes $L$ and $L-\delta$, respectively, their values of $\gamma$, Eq.~(\ref{gamma}), must remain identical. This implies
\begin{equation}
  g_2(L-\delta) = g_2(L)\frac{k_F}{k_F+q/2}.
\end{equation}
To lowest order in perturbation theory, the energy of the excitation can be extracted from Eq.~(\ref{perten}). The frequency of the excitation reads
\begin{equation}
  \hbar\omega(q) = E(L-\delta)-E(L)=\frac{\hbar^2k_F}{m}\left(1-\frac{4}{\gamma}\right)q,
\end{equation}
in perfect agreement with the known lowest order result \cite{Cazalilla}.

We can generalise the weak-coupling results to all orders of the EFT potential. Define the formal series
\begin{equation}
  V(q)=\sum_{n=1}^{\infty}g_{2n}q^{2n}.
\end{equation}
In order to keep all dimensionless constants invariant under the size shrinking, it is easy to see that the bare coupling constants must satisfy
\begin{equation}
  ng_{2n}(L-\delta) = g_{2n}(L)\frac{k_F}{k_F+q/2}.
\end{equation}
The excitation spectrum then has the following frequency
\begin{equation}
  \hbar \omega(q) = \left[\frac{\hbar^2k_F}{m}-\frac{1}{2\pi}\sum_{n=1}^{\infty}(2k_F)^{2n}g_{2n}\right]q=\left[\frac{\hbar^2k_F}{m}+\frac{1}{2\pi}(V(0)-V(2k_F))\right]q
\end{equation}
The above result coincides with the weak-coupling result of the previous section (i.e. by expanding the speed of sound to lowest order and using the weak-coupling value of $g_0$).

We can now study the method non-perturbatively. The main idea is that the excitation frequency is given, to lowest order in $q$ by
\begin{equation}
  \hbar\omega(q) = E(L-\delta)-E(L)\approx -E'(L)\delta.\label{ex}
\end{equation}
It is {\it very important} to remember that the derivative $E'(L)$ must be done while keeping the dimensionless coupling constants unchanged.
We can parametrise the energy of the two-body system as $E(L)=\hbar^2k^2/m$, where $k$ is the solution to the following equation
\begin{equation}
  k=k_F-\frac{k_F}{\pi}\theta_k,
\end{equation}
where $\theta_k$ is the two-fermion scattering phase shift. In terms of $k$, the expression for the excitation spectrum (\ref{ex}) can be rewritten as $\hbar\omega(q)=\hbar v q$, with
\begin{equation}
  \hbar v = \frac{\hbar^2}{m}k\frac{\mathrm{d}k}{\mathrm{d}k_F},
\end{equation}
where we have used that $\delta=L^2q/4\pi$.

\subsection{Flat bands}
A flat band is nothing but a single particle dispersion, on a lattice, that is infinitely degenerate. That is, a band $\sigma$ is flat if its energy is a constant
\begin{equation}
\epsilon_{\sigma}(q)=E_\sigma, \hspace{0.3cm} \forall q \in \mathrm{BZ}.
\end{equation}
For simplicity, we will assume the trivial case of a system with a single flat band, but note that the results that follow apply when there are other, dispersive bands \cite{ValienteZinnerFlat}. We set a single particle on the flat band interacting with a probe impurity, and study the scattering states. The Lippmann-Schwinger equation takes the form ($z=E_{\sigma}+i\eta$)
\begin{equation}
\bra{\mathbf{k}'}T(z)\ket{\mathbf{k}}=\bra{\mathbf{k}'}V\ket{\mathbf{k}}+\frac{1}{\Omega}\int_{\mathbf{BZ}}\mathrm{d}\mathbf{q}\frac{\bra{\mathbf{k}'}V\ket{\mathbf{q}}\bra{\mathbf{q}}T(z)\ket{\mathbf{k}}}{i\eta}.\label{Tflat1}
\end{equation}
We immediately observe the problem of having $i\eta$ in the denominator ($\eta\to 0^+$). Fortunately, we can solve this by rewriting the $T$-matrix as
\begin{equation}
\bra{\mathbf{k}'}T(z)\ket{\mathbf{k}}\equiv i\eta \bra{\mathbf{k}'}t(z)\ket{\mathbf{k}}.\label{Tflat2}
\end{equation}
Eq. (\ref{Tflat1}) now reads
\begin{equation}
\bra{\mathbf{k}'}T(z)\ket{\mathbf{k}}=\bra{\mathbf{k}'}V\ket{\mathbf{k}}+\frac{1}{\Omega}\int_{\mathbf{BZ}}\mathrm{d}\mathbf{q}\bra{\mathbf{k}'}V\ket{\mathbf{q}}\bra{\mathbf{q}}t(z)\ket{\mathbf{k}}.\label{Tflat3}
\end{equation}
Using the definition (\ref{Tflat2}) in Eq.~(\ref{Tflat3}) and taking the limit $\eta\to 0^+$, we obtain
\begin{equation}
-\bra{\mathbf{k}'}V\ket{\mathbf{k}}=\frac{1}{\Omega}\int_{\mathrm{BZ}}\mathrm{d}\mathbf{q} \bra{\mathbf{k}'}V\ket{\mathbf{q}}\bra{\mathbf{q}}t(z)\ket{\mathbf{k}}.
\end{equation}
The above equation can be written in operator form as
\begin{equation}
-V=Vt(z).\label{Vt}
\end{equation}
Let us solve the above equation by first diagonalising $V$. Its eigenfunctions will be called $\ket{\alpha_{\mathbf{k}}}$ and their associated eigenvalues are $V(\alpha_{\mathbf{k}})$.  Eq.~(\ref{Vt}) takes the matrix form
\begin{equation}
-\Omega V(\alpha_{\mathbf{q}})\delta(\mathbf{q}-\mathbf{q}')=V(\alpha_{\mathbf{q}'})\bra{\alpha_{\mathbf{q}'}}t(z_0)\ket{\alpha_{\mathbf{q}}}.
\end{equation}
The above equation allows us to set 
\begin{equation}
\bra{\alpha_{\mathbf{q}'}}t(z_0)\ket{\alpha_{\mathbf{q}}}=-Omega\delta(\mathbf{q}-\mathbf{q}'),
\end{equation}
The scattered wave $\ket{\psi}_s=G_0(z)T(z)\ket{\mathbf{k}}=\bra{\mathbf{k}'}t(z)\ket{\mathbf{k}}$.  It is easy to see that the full scattering state takes the simple form
\begin{equation}
\ket{\psi}=\hat{P}_{I_0}\ket{\mathbf{k}},
\end{equation}
where $\hat{P}_{I_0}$ the projection onto the kernel of $V$ ( in the flat-band subspace, when more than one band is present).

\section{Hardcore bosons on a lattice: weakly or strongly interacting?}
We move back now to consider a system where indirect methods can provide answers to a seemingly complicated problem. Let us take a many-body system of bosons on a three-dimensional cubic lattice interacting via a hard core zero-range interaction and nearest-neighbour tunnelling with rate $J$. This system is equivalent to the XY spin model in the same geometry, where each boson corresponds to a magnon on top of the totally ferromagnetic state $\ket{\uparrow\uparrow\ldots\uparrow}$. In order to be able to make any analytical predictions, we should work near the continuum limit, that is, at very low filling factors $\nu=N/N_s$, where $N$ is the number of bosons and $N_s$ the number of lattice sites. If the lattice spacing is denoted by $d$, then the continuum density of the system is given by $\rho=\nu/d^3$. We can use Bogoliubov theory if the system is dilute, i.e. if $\rho a^3 \ll 1$, where $a$ is the two-body scattering length. Hence, the question is whether these strongly interacting bosons have a small or a large scattering length, and how it scales with the only natural length of the system, the lattice spacing $d$.

Firstly, we reintroduce the interaction strength $U$, which is taken to infinity at the end of the calculation. At zero energy, and on the energy shell, the T-matrix is a constant $t$, given by
\begin{equation}
 t=\frac{1}{1/U-W(0)}, 
 \end{equation}
where
\begin{equation}
W(E)=\frac{1}{(2\pi)^3}\int_{\mathrm{BZ}}d\mathbf{q}\frac{1}{E-\epsilon(\mathbf{q})+i\eta},
\end{equation}
where BZ is the first Brillouin zone $[-\pi,\pi)^3$ and 
\begin{equation}
\epsilon_0(\mathbf{q})=-4J\sum_{\alpha=x,y,z}\left[\cos(k_{\alpha}d)-1\right].
\end{equation}
Therefore, in the hardcore limit, we obtain $t=-1/W(0)$. Using $W(0)=-0.252733(m/\hbar^2d)$, we obtain the scattering length 
\begin{equation}
a=0.314868d. 
\end{equation}
The above analysis shows that, near the continuum limit, where the dispersion is nearly quadratic, the scattering length can be made arbitrarily small as the lattice spacing $d$ is reduced. For fixed $d$ and fixed $\nu\ll 1$, we may used Bogoliubov theory to describe the strongly interacting lattice gas, since, even though it appears very strongly interacting, it is effectively very weakly interacting! 

\section{Conclusions}
We have presented a number of methods pertaining to few-body systems that can be used in the many-body problem. In particular, two-body methods have been used, which are simplest to understand and implement, and are most developed so far. The methodology here presented, though not complete, tries to put forward ideas whose power has only started to be recognised recently. Although we have divided the methods into indirect and direct, these complement each other and the line separating both types of methods is not strict. The main conclusion that can be drawn from these methods is that strongly interacting systems, that may look formidable at first, may actually be treated in some instances by simple theoretical methods if the few-body sector is under control. This new trend is now in its infancy, and further and more exciting developments are still to come.

\end{document}